\documentclass[mathleft
]{an}
\usepackage{graphicx}
\usepackage{times}
\usepackage[switch]{lineno}

\begin{document}
\linenumbers
\modulolinenumbers[5]

\Pagespan{789}{}
\Yearpublication{2010}%
\Yearsubmission{2010}%
\Month{11}%
\Volume{999}%
\Issue{88}%

\title{Constraining the core-rotation rate in red-giant stars \\
from Kepler space photometry}

\author{P.\,G. Beck\inst{1}\fnmsep\thanks{Corresponding author: \email{paul.beck@ster.kuleuven.be}\newline}\and
J. De Ridder\inst{1} \and  
C. Aerts\inst{1,2} \and
T. Kallinger \inst{1} \and
S. Hekker\inst{3,4} \and  \newline
R.\,A. Garcia\inst{5} \and  
B. Mosser\inst{6} \and  
G.\,R. Davies\inst{5}
}
\titlerunning{Constraining the core-rotation rate in red giant stars 
from Kepler space photometry }
\authorrunning{P.\,G. Beck et\,al.\ }
\institute{
Instituut voor Sterrenkunde, Universiteit Leuven, Celestijnenlaan 200 D, Bus 2401, 3001 Heverlee, Belgium
\and Department of Astrophysics, IMAPP, Radboud University Nijmegen, PO Box 9010, 6500 GL Nijmegen, NL.
\and Astronomical Institute ÕAnton PannekoekÕ, University of Amsterdam, Science Park 904, 1098 XH Amsterdam, 
NL.
\and School of Physics and Astronomy, University of Birmingham, Edgbaston, Birmingham B15 2TT, UK.
\and
Laboratoire AIM, CEA/DSM Ð CNRS - Universit\'e Denis Diderot Ð IRFU/SAp, 91191 Gif-sur-Yvette Cedex, France
\and
LESIA, CNRS, Universit\'e Pierre et Marie Curie, Universit\'e Denis Diderot,
Observatoire de Paris, 92195 Meudon cedex, France}
\received{15 August 2012}
\accepted{accepted}
\publonline{later}

\keywords{stars: oscillations, rotation -- stars: individual KIC\,8366239, KIC\,5356201, KIC\,12008916}

\abstract{%
Rotation plays a key role in stellar structure and its evolution. Through transport processes which induce rotational mixing of chemical species and the redistribution of angular momentum, internal stellar rotation influences the evolutionary tracks in the Hertzsprung-Russell diagram. In turn, evolution influences the rotational properties. Therefore, information on the rotational properties of the deep interior would help to better understand the stellar evolution.
However, as the internal rotational profile cannot be measured directly, it remains a major unknown leaving this important aspect of models unconstrained. We can test for nonrigid rotation inside the stars with asteroseismology. Through the effect of rotational splitting of non-radial oscillation modes, we investigate the internal rotation profile indirectly. Red giants have very slow rotation rates leading to a rotational  
splitting on the level of a few
tenth of a $\mu$Hz. Only from more than 1.5 years of consecutive data from the NASA \textit{Kepler} space telescope, these tiny variations could be resolved. A qualitative comparison to theoretical models allowed constraining the core-to-surface rotation rate for some of these evolved stars. In this paper, we report on the first results of a large sample study of splitting of individual dipole modes.}

\maketitle

\section{Introduction}
\sloppy
Towards the end of the core hydrogen burning phase, stars like the Sun experience an exhaustion of hydrogen in the core. The contraction of the remaining helium core and re-ignition of hydrogen in a shell around it, give rise to an expansion of the convective envelope. This leads to a drop in the effective temperature but to an increase in luminosity. Such stars of spectral types G and K are known as red giants.

In the last decade, the existence of high radial order pressure-mode oscillations in evolved stars could be confirmed with ground based state-of-the-art spectrographs \hbox{capable} of a meters-per-second precision measurements, e.g. $\xi$\,Hya with \textsc{Coralie} (Frandsen et\,al.\ 2002) and with satellite telescopes such as \textsc{Most}  (Walker et\,al.\ 2003), \textsc{CoRoT} (Baglin et\,al.\ 2006) and \textit{Kepler}  (Borucki et\,al.\ 2010) sensitive to variations down to a few parts-per-million (ppm). 
The firm detection of non-radial modes in multiple red giants, as demonstrated by De Ridder et\,al.\ (2009), allowed the succesful application of asteroseismic techniques to investigate the structure of these evolved stars.  These solar-like oscillations are stochastically excited by convective motions in the outer layers. In the pre-space era only pure pressure modes were resolved in red giants, which propagate through outer layers of these stars.  Hence, exploiting such modes can only help to constrain the parameters of the outer convective envelope, leaving asteroseismologists blind to the conditions in the central region of the stars. 
\begin{figure*}[ht!]
\centering
\includegraphics[width=\textwidth,height=60mm]{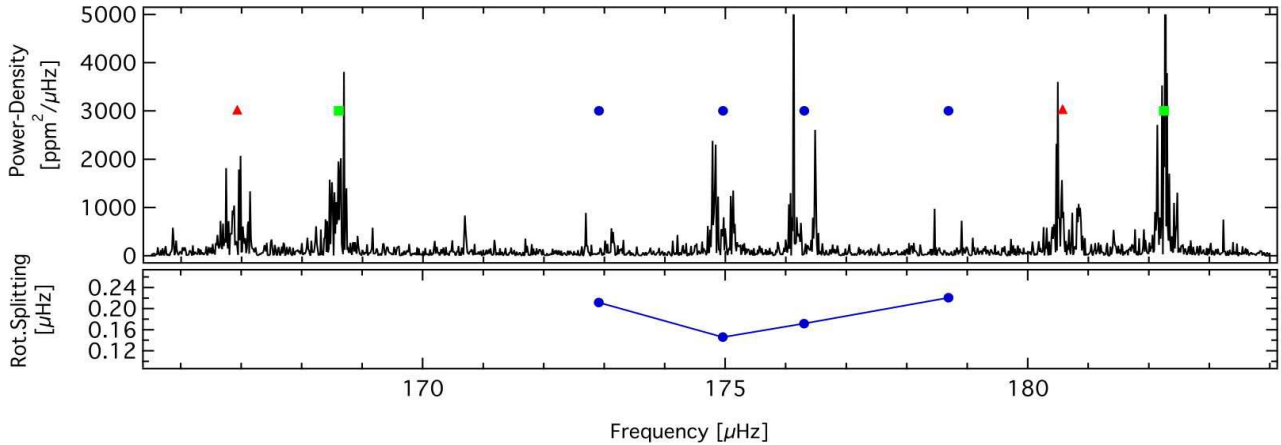}
\caption{The power density spectrum in the upper panel shows the central radial order in the oscillation power excess of KIC\,8366239. Extracted modes of the spherical degrees $\ell$ = 0 and 2 are shown with green squares and red triangles, respectively. The centre component ($m=0$)  of rotationally split dipole modes is shown with blue dots. The lower panel shows the value of the rotational splitting of the individual dipole modes.}
\label{fig:kic8366}
\end{figure*}

With the advent of the space mission \textit{Kepler}, oscillation modes with a substantial degree of mixed character between pressure and gravity modes were firmly identified in the oscillation spectra of red-giant stars by Beck et\,al.\ (2011) and further exploited by  Bedding et\,al.\ (2011) and Mosser et\,al.\ (2011). 
These mixed modes not only sound the outer regions, but also probe the core of the red giants. Therefore, they provide asteroseismic information on the overall density gradient which depends on the evolutionary stage. Bedding et\,al.\ (2011) and Mosser et\,al.\ (2011) therefore could discriminate between H-shell and He-core burning red giants. From the analysis of such dipole mixed modes, Beck et\,al.\ (2012) have shown that non-rigid rotation is in fact present in the interior of these stars. Radial differential rotation has also been detected by Deheuvels et\,al.\ (2012) in a young giant at the beginning of the RGB phase.

\subsection{\textit{Kepler} Data}
The \textit{Kepler} space telescope, which became operational in early 2009, collects continuous high-precision data of more than 160\,000 stars in a broad photometric passband. For this study, we use $\sim$880 days of 
photometric data (Q0-Q10), taken in the long-cadence mode, i.e. a photometric measurement approximately integrated for 30 minutes, and processed following Garcia et\,al.\ (2011). Therefore, we can only study those stars whose maximum frequency of the oscillation power excess $\nu_{\rm max}$ is located below the Nyquist frequency of 283 $\mu$Hz. 

Approximately every three months, new data are released which prolong the total length of the time series by another 90 days and improve the frequency resolution of the power-density spectra. 
As length of these high-quality consecutive photometric datasets increases, we are able to study finer details and the long-lived g-dominated mixed modes in red giants.

\subsection{The internal rotation profile of stars on the red giant branch}
Rotation has a substantial impact on the result of stellar evolution (e.g., Eggenberger et\,al.\ 2010). It induces mixing of chemical species and also the transport of angular momentum. Consequently rotation  influences the evolution beyond the main sequence and modifies the evolutionary tracks in the Hertzsprung-Russell Diagram (HRD). 

In general one expects that the rotational profile of a red giant is governed by a rapidly rotating helium core in a slowly rotating convective envelope. In this envelope, nonrigid rotation along the radial direction is expected due to the local conservation of angular momentum during the expansion phase. The first to note the necessity of such an internal rotation profile were Sweigart \& Mengel (1979). 
From the analysis of the observed surface abundance ratios of chemical species in red giants, they found that the required angular velocities of the main-sequence progenitors would be prohibitively large, 
if solid body rotation is assumed in the red giant phase.
Until recently, the only direct observational proof of rotation in red giants concerns their projected surface rotation. Large sample studies of ground based spectroscopy 
(e.g., Hekker \& Mel\'endez 2007) have revealed that typical surfaces are indeed rotating slowly.

The internal rotational profile and even more the core rotation can not be monitored with any direct observational techniques or classical stellar parameter as the gaseous envelope blocks the view onto the deeper layers. However, we can learn more about rotation from the analysis of nonradial modes. 
Rotation breaks the degeneracy of a mode with the spherical degree $\ell>0$ and generates a $(2\cdot\ell+1)$-tuplet of modes. 
Each mode with an azimuthal order $m$$\neq$0 will be shifted away from the zonal mode ($m$=0). This frequency separation,
\begin{equation}
\delta f_{n,\ell,m} = \Delta m\cdot\frac{\Omega}{2\pi}\cdot(1-C_{nl}),
\end{equation}
depends on the averaged rotation rate $\Omega$ of the cavity in which a given mode propagates and the Ledoux constant $C_{nl}$ (Ledoux 1951). 
For pure dipole ($\ell=1$) pressure modes (p modes) of high radial order, $C_{nl}$ is approximately 0, while for pure gravity dipole modes (g modes) of high radial order the Ledoux constant equals 0.5. 

The amplitudes of the individual components of rotationally split multiplets are highly dependent on the inclination of the star's rotation axis towards the observer, which in principle allows us to derive the inclination of the star (Gizon \& Solanki 2003, Ballot et\,al.\ 2006). For a recent review of the phenomenon of rotational splitting, we refer to Aerts et\,al.\ (2010). 

Dipole mixed modes, as we find them in the power spectra of red giants, share the pressure radial order they are located in, but have different degrees of mixing between pressure (p) and gravity (g) modes. Therefore the value of $C_{nl}$ has to be determined from models and will range between 0 and 0.5. 
Observationally, we can also discriminate between the p- and g-dominated mixed modes from their position of the modes in the power spectrum and the mode line width. The closer a mixed dipole mode is located to the position of the pure pressure dipole mode, the higher is its p-mode component (Mosser et\,al. 2012a, and references therein). Another parameter, which varies with the degree of mixing is the mode lifetime. Due to their low mode inertia, p-dominated mixed modes are short lived and exhibit broad peaks in the power spectrum, while long-lived g-dominated modes exhibit sharp, unresolved mode peaks due to their high mode inertia.

An analysis of the rotational splitting of dipole modes was presented by Beck et\,al.\ (2012). For the star KIC\,8366239 they found that the splitting of the g-dominated dipole modes is on average 1.5 times larger than the splitting of the p-dominated modes. 
Figure \ref{fig:kic8366} shows the mixed dipole modes in the central radial order of the oscillation power excess of KIC\,8366239 and their measured rotational splitting.
Comparing the derived ratio of splitting of g- and p-dominated modes with theoretically computed splittings Beck et\,al.\ (2012) interpreted the larger g-dominated mode splitting as the signature of the fast rotating core and concluded that the core must be rotating at least 10 times faster than the surface of the star.  

\section{Towards larger samples}
\begin{figure}[t!]
\centering
\includegraphics[width=0.47\textwidth]{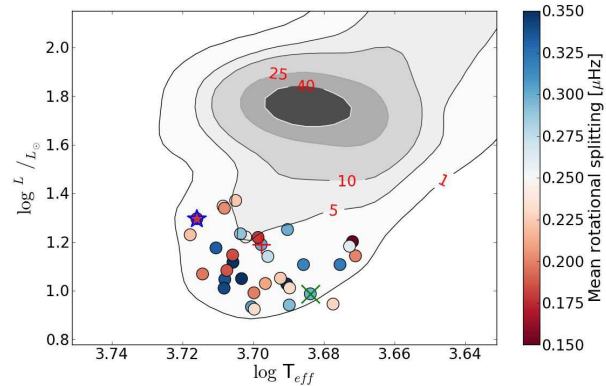}
\caption{This Hertzsprung-Russell diagram compares the mean rotational splitting of 40 stars (colored dots) which are in the same evolutionary state as KIC\,8366239 (blue $\star$), KIC\,5356201 \hbox{(green x),} KIC\,12008916 (red $+$). 
The color of the dots represents the mean value of the rotational splitting found in these stars.
The contour surfaces reflect the density distribution of the best 1000 red giants in the sample. The darkest areas mark approximately the position of the densely populated red clump. Numbers in red indicate the star count per bin, for which the contour surfaces have been drawn. 
} \label{fig:splitHRD}
\end{figure}

Rotation is detectable in many red giant stars. Following the result of the core rotation, we are  investigating large samples of red-giant stars showing rotational splitting, to understand how stellar evolution modifies rotational properties as a star advances towards the helium-core burning phase. We therefore tested if the stars discussed in Beck et\,al.\ (2012) are compatible with the rotational properties of stars on the lower red giant branch. We analysed the well studied group of 1400 red giants observed by \textit{Kelper} (e.g. Hekker et\,al.\ 2010, Kallinger et\,al.\ 2011, Mosser et\,al.\ 2012a, etc.). Figure \ref{fig:splitHRD} shows their distribution in the HRD. 

At this stage we limited our sample to stars which exhibit their power excess $\nu_{\rm max}$ at frequencies below the Nyquist frequency but higher than 100$\mu$Hz. 
Within this range we found some 40 stars which show rotational splitting. These stars are shown in figure \ref{fig:splitHRD}. All stars belong to the class of low-luminosity red giants (Bedding et\,al.\ 2010) and have stellar radii of approximately 3.5 to 6.5 R$_\odot$. The observed period spacings for these stars, were determined from the center component of the triplets ($m=0$) and therefore are independent of rotational splitting. This parameter indicates that all stars in the sample are in the hydrogen-shell burning phase (Bedding et\,al.\ 2011, Mosser et\,al.\ 2011).

While KIC\,8366239 has an average splitting of 0.18$\mu$Hz, the two other stars have an average splitting of 0.30 $\mu$Hz. As we can see from figure \ref{fig:splitHRD}, these values are in the middle of the range of mean rotational splitting, which we found in other low-luminosity red giant stars.  As it can be seen from the color coding of this figure the rotational splitting is decreasing, as a star ascends the RGB. This decrease is governed by the increasing stellar radii (Mosser et\,al.\ 2012b).

The non-detection of rotation in numerous stars, based on the present dataset does not necessarily exclude the presence of rotation. Although the  rotational frequency measured through rotational splitting is independent of the inclination, the amplitude ratios of the split components are not. On one hand, this enables us to constrain the inclination of the rotation axis towards us. On the other hand, this dependency makes it hard to detect rotation if the inclination of the rotation axis is close to 0$^\circ$, as the amplitudes of the $m=\pm$1 components of the split dipole mode disappear in the background noise. For some stars higher up the RGB the effects of rotation on nonradial modes might be too small to be resolved with the present data and need longer time series. Another difficulty arises if the size of the rotational splitting is on the order of the period spacing. This leads to closely neighboring, even overlapping frequency peaks, which give an underestimation of the rotational splittings. In this instance, applying the asymptotic relation for mixed modes
as described by Mosser et\,al.\ (2012b) enables us to identify such stars and extract the correct value of the rotational splitting.

\subsection{Comparison with ground based spectroscopy}
To obtain independent estimates of the stellar fundamental parameters, we have taken ground based high resolution spectra of numerous red giants with the \textsc{Hermes}-spectrograph (Raskin et\,al.\ 2011) mounted on the 1.2 meter \textsc{Mercator} telescope on the Canary Islands, Spain, which are currently under investigation.

Comparing the results on rotation from rotational splitting with those derived from spectroscopy was found to be difficult. Beck et\,al.\ (2012) have already obtained spectroscopy for two of the stars and noted a difference between the asteroseismically determined internal rotation rate from the most pressure-like mode (i.e. dominated by the rotation of the envelope) and the projected surface velocity $v\sin i$. This offset could not be explained by the inclination, which is known for the stars from geometry of the rotationally split multiplets. From the analysis of the rotational kernels it was shown that even pressure-dominated mixed modes have a substantial contribution of about 30\% from the core rotation. This leads to a substantially higher rotation rate derived from splitting than estimated from the  $v\sin i$. This result has been confirmed by Mosser et\,al.\ (2012b) who have shown that the mean rotational splitting of dipole modes is a good proxy of the core rotation.

\section{What can we learn from the core rotation of stars?}
For several decades, the internal rotational gradient of stars remained unconstrained and hence a free parameter in stellar models. 
Together with similar results for young massive B stars (3-5 times the rotation rate of the surface, Aerts et\,al.\ 2003, Pamyatnykh et\,al.\ 2004), white dwarfs (Kawaler et\,al.\ 1999) and also recently for subgiant stars (6-10 times, Deheuvels et\,al.\ 2012), the determination of the core-rotation rate in red-giant stars sheds light on the core-rotation rate as a function of the stellar evolution.

The exchange of results and ideas at the ESF meeting in Obergurgl, Austria has also shown, that the experimental and theoretical results do not agree (Eggenberger et\,al.\ 2012; Ceillier et\,al.\ 2012)  as it is also the case when comparing the inferred differential radial rotation of the Sun (e.g. Garc\'\i a et\,al.\ 2007) with the modeled ones (e.g. Pinsonneault et\,al.\ 1989; Turck-Chi\`eze et\,al.\ 2010). 
In evolved red giants, the theoretical results by Marques et\,al.\ (2012) predict a core-to-surface rotation rate which is too high by a factor of 10 up to 100, compared to the observations, when no angular momentum is transferred from the core to the surface.

\acknowledgements 
We would like to thank the European Science Foundation (ESF) for the organization of the conference and financial support and Dr. Markus Roth for the interesting scientific program of this very fruitful conference.

We acknowledge the work of the team behind Kepler. Funding for the Kepler Mission is provided by NASAÕs Science Mission Directorate. The research leading to these results has received funding from the European Research Council under the European Community's Seventh Framework Programme (FP7/2007--2013)/ERC grant agreement n$^\circ$227224 (PROSPERITY), as well as from the Research Council of K.U.Leuven grant agreement GOA/2008/04.
 J.D.R. and T.K. were supported by the Fund for Scientific Research. S.H. was supported by the Netherlands Organisation for Scientific Research (NWO).

\end{document}